\documentclass[aps,prl,twocolumn,superscriptaddress,amsmath,amssymb]{revtex4-1} 
\usepackage{graphicx}  
\usepackage{dcolumn}   
\usepackage{comment}
\usepackage{bm}        
\usepackage{verbatim}
\usepackage{siunitx}   
\usepackage{xcolor}
\usepackage{hyperref}
\usepackage[sort&compress]{natbib} 		   

\expandafter\def\expandafter\normalsize\expandafter{%
	\normalsize
	\setlength\abovedisplayskip{15pt}
	\setlength\belowdisplayskip{15pt}
	\setlength\abovedisplayshortskip{15pt}
	\setlength\belowdisplayshortskip{15pt}
      }

\newcommand{\kBT}{{k_\te B T}}
\newcommand{\kB}{{k_\te B}}
\newcommand{\Dr}{{D_\te r}}
\newcommand{\DR}{{D_\te r}}
\newcommand{\Deff}{{D_\te{eff}}}

\newcommand{\Ra}{{\rho_\te a}}
\newcommand{\Rp}{{\rho_\te p}}

\newcommand{\la}{{\lambda_{\te a}}}
\newcommand{\lp}{{\lambda_{\te p}}}

\newcommand{\te}[1]{\mathrm{#1}}

\newcommand{\df}{\mathrm{d}}
\newcommand{\vv}[1]{\mathbf{#1}}
\begin{document}

\newcommand\uleipE{\affiliation{
Peter Debye Institute for Soft Matter Physics, 
Leipzig University, 04103 Leipzig, Germany}}
\newcommand\uleipT{\affiliation{
Institute for Theoretical Physics, 
Leipzig University, 
04103 Leipzig, Germany}}
\newcommand\chau{\affiliation{ 
 Charles University,  
 Faculty of Mathematics and Physics, 
 V Hole{\v s}ovi{\v c}k{\' a}ch 2, 
 CZ-180~00~Praha, Czech Republic 
}}

\title{Active-Particle Polarization Without Alignment Forces}

\author{Nicola Andreas S\"oker}\email{nicola.soeker@.uni-leipzig.de}\uleipE
\author{Sven Auschra}\email{sven.auschra@itp.uni-leipzig.de}\uleipT
\author{Viktor Holubec}\email{viktor.holubec@mff.cuni.cz}\uleipT\chau
\author{Klaus Kroy}\email{klaus.kroy@uni-leipzig.de} \uleipT
\author{Frank Cichos}\email{frank.cichos@uni-leipzig.de}\uleipE
\date{\today}
\begin{abstract}
Active-particle suspensions exhibit distinct polarization-density patterns in activity landscapes, even without anisotropic particle interactions. Such polarization without alignment forces is at work in motility-induced phase separation and betrays intrinsic microscopic activity to mesoscale observers. Using stable long-term confinement of a single hot microswimmer in a dedicated force-free particle trap, we examine the polarized interfacial layer at a motility step and confirm that it does not exert pressure onto the bulk. Our observations are quantitatively explained by an analytical theory that can also guide the analysis of more complex geometries and many-body effects.
\newline
 \end{abstract}
\maketitle


Active matter can succinctly be characterized as matter made from ``animalcules'', a type of non-equilibrium molecules \cite{Lane2015Animalcules}. As a consequence, it can exhibit unusual material properties that would be strictly forbidden in conventional materials by symmetries implicit in the condition of (local) thermal equilibrium.  Most distinguished examples are found within living active matter. They stimulate the development of mathematically tractable synthetic toy models that can closely mimic distinctive traits of life, thereby providing good grounds for saying that ``biology becomes physics'' \cite{Cates2012DiffusivePhysics,Contera2019Nano}.

On a fundamental level, the hallmark of living and synthetic active matter is its local, particle-level entropy production. However, in practice, its functionally relevant component is commonly dwarfed by so-called ``housekeeping'' fluxes that abound in non-equilibrium environments, hampering its detection. 
Persistent ring currents, signaling said entropy production \cite{Gnesotto2018BrokenDetailedBalance}, and mesoscopic action-reaction symmetry breaking, signaling hidden currents \cite{Steffenoni2016InteractingBath,Basu2015StatForces}, have recently been proposed as more tangible mesoscopic signatures of the local non-equilibrium dynamics characteristic of active matter. But they are not necessary traits.
Other emergent mesoscale signatures of activity suffer from the opposite limitation. For example, the phenomenon of motility-induced phase separation~\cite{Solon2018GeneralizedMatter,Cates2015MIPS,Cates2013WhenSeparation} in active-particle suspensions
is arguably closely related to inelastic collapse in granular flows~\cite{Shida1989InelasticCollisions, McNamara1992InelasticCollapse}. It is indicative of dissipative interactions but not, \emph{per se}, of activity on the particle level. Moreover, a very similar phenomenology is predicted for ordinary colloidal particles that differ (only) in their diffusivities~\cite{Schnitzer1993TheoryChemotaxis,Weber2016BinaryMixturesDemix}. Similarly, the boundary-layer formation at (effectively) inelastic walls \cite{Elgeti2013WallAccumulation,Speck2016IdealBulkPressure,Wagner2017ABPsUnderConfinement}, often underlying ratcheting \cite{Seifert2012MolecularMachines,Holubec2017ThermalRatchetConfGeom}, only requires persistent paths near curved boundaries, as in the spontaneous cortex formation in confined semiflexible polymer solutions \cite{Lieleg2007BundlesSemiFlexPolymers}, say.

In this Letter and a companion article \cite{Auschra2020ActivityFieldsLong}, we investigate the co-localization of density modulations and polarization, which has so far received less attention as a robust alternative mesoscopic criterion for detecting particle-level activity. 
Our experiments exploit our precise control over autonomous Janus-type microswimmers via photon nudging \cite{Qian2013HarnessingNudgingb,Bregulla2014StochasticNudging,Selmke2018TheoryTransport,Selmke2018TheoryConfinement}. 
Notably, apart from weak gravitational forces and interactions with the container walls, which are largely irrelevant to our experiment, our setup is entirely force-free. A two-dimensional rectangular arena is divided into two regions of diverse activity, separated by a sharp activity step, as found as a concomitant feature of most actual physical boundaries. Here, it is however realized without inserting any physical wall, as otherwise often done in experiments and computer simulations. We thereby avoid possible unintended (hydrodynamic, electrostatic, steric,\dots) side effects, which could uncontrollably alter the particle density, current, and orientation. This allows us to experimentally confirm, on the single-particle level, that the interfacial polarization is emerging from hidden bulk currents and imbalances \cite{hermann2020PolStateFct} rather than ``controlling'' the bulk states \cite{Solon2018GeneralizedMatter}. Our data reveal rich mutual relations between the particle density, polarization, and motility, which are all quantitatively explained by an analytical theory~\cite{Auschra2020ActivityFieldsLong} that goes beyond previous literature results \cite{Schnitzer1993TheoryChemotaxis,Malakar2018RunAndTumble1D,Sharma2017BrownianActivity,Fischer2020Quorum-sensingMotility,fischer2020erratum}. 


\emph{Materials and Methods:} 
Janus microswimmers are constructed of a $\SI{1.5}{\micro\meter}$ diameter polystyrene (PS) core (microParticles GmbH) and a $\SI{50}{\nano\meter}$ thin gold hemisphere. The particles are propelled by optically controlled self-thermophoresis \cite{Qian2013HarnessingNudgingb,Bregulla2014StochasticNudging,Selmke2018TheoryTransport,Selmke2018TheoryConfinement}. The sample consists of a $\SI{2.4}{\micro\meter}$ thin water film confined by two microscopy cover slips coated with Pluronic F127 to prevent particle adsorption, and sealed with polydimethylsiloxane to prohibit evaporation.
The particle's in-plane motion was observed in an inverted microscope Olympus (IX-73) under darkfield illumination (Olympus DF condenser) using $\SI{1}{\milli\second}$ duration LED flashes (Thorlabs SOLIS-3C). The illumination was synchronized with a CCD camera (Hamamatsu Orca-Flash4.0 V2) at an inverse frame rate of $\SI{20}{\milli\second}$. The activity of the particle was adjusted by a $\lambda=\SI{532}{\nano\meter}$ laser with a homogeneous intensity over the whole sample area, whose light intensity could be varied by an Acousto-Optical Modulator (Isomet 1260C) between 0 and $\SI{12}{\micro\watt\micro\meter^{-2}}$. The real-time processing of particle position and orientation as well as the feedback control was implemented in LabVIEW as described in Ref.~\cite{Bregulla2014StochasticNudging}. The setup's overall feedback latency time amounts to $\SI{10}{\milli\second}$, which is negligible for the experimental results.

\begin{figure}[tb!]
    \centering
    \includegraphics[height=0.9\columnwidth]{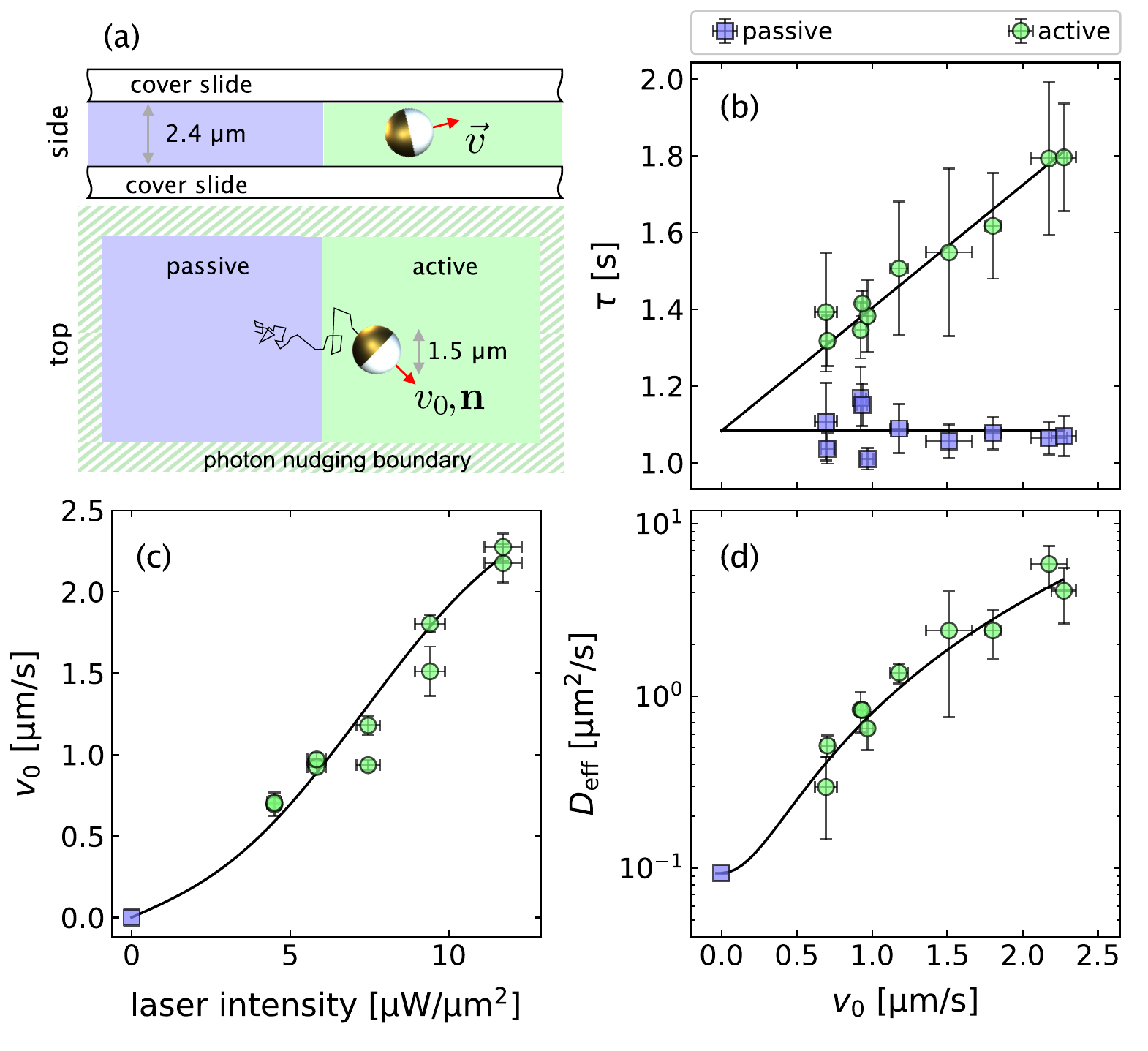}
    \caption{Setup and parameter measurement.(a) Sketch of the rectangular arena, in which a  $\SI{1.5}{\micro\meter}$ Janus particle is confined by photon nudging. Position and in-plane orientation $\vv n$ are observed in  darkfield microscopy. 
    (b) Orientational correlation time $\tau$ for the active/passive bulk region as a function of the in-plane propulsion speed $v_0$. The horizontal and linear fits serve to identify $\tau_\te{p,a}$. (c) In-plane propulsion speed $v_0$ as a function of the incident laser intensity, with a fit accounting for the weakly variable particle-wall alignment. (d) Effective diffusion coefficient $\Deff$ in the active bulk, obtained from the slopes of the late-time MSD, and  Eq.~\eqref{eq:deff} (line) using $\tau_\te{a}$ from (b). In (b)-(d)  squares and circles correspond to the passive and active region, respectively, and error bars show 95 \% confidence intervals for Gaussian error propagation.
}
    \label{fig:figure1}
\end{figure}

The feedback controlled photon nudging \cite{Qian2013HarnessingNudgingb} allowed us to define a virtual rectangular arena of $\SI{6}{\micro\meter}$ $\times$ $\SI{5}{\micro\meter}$ for the swimmer (Fig.~\ref{fig:figure1}(a)). Outside the arena, the heating laser is turned on only if the swim direction of the Janus particle points towards the arena, yielding a mean propulsion speed of $\SI{2.5}{\micro\m\per\second}$. We further divided the central area of the arena into a passive and an active region,
in which the laser was always off and on, respectively, irrespective of the particle orientation. Thereby, an activity landscape with a step profile $v(x) = v_0 \Theta(x)$ of the particle velocity is realized. The step function $\Theta(x)$ is $0$ in the passive (p, $x<0$) and $1$ in active (a, $x>0$) region. The experiments were performed with a single Janus particle, with observation times of about 2 days for each of the laser powers depicted in Fig.~\ref{fig:figure1}(c). 

The particle's in-plane translational and rotational motion is characterized by its  translational and rotational diffusion coefficients $D$ and $D_r$. They are estimated from the characteristic time scales exhibited by the translational mean-squared displacement (MSD) and the in-plane orientational auto-correlation function (ACF), in either the active or passive region.
Long trajectories that live solely inside the passive/active region cannot be obtained due to the limited size of the active/passive region. The particle frequently commutes between the two regions, leading to many short trajectories that are concatenated in such a way that the in-plane and out-of-plane angles of the Janus particle matched.

In both regions, the value of the diffusion coefficient $D= \SI{0.094}{\micro\meter^2\second^{-1}}  \pm \SI{0.002}{\micro\meter^2\second^{-1}}$  is roughly 1/3 of the expected Stokes-Einstein value $\kB T/6\pi\eta R$ in the bulk, which we ascribe to hydrodynamic interactions of the swimmer with the confining cover slips \cite{Happel1981LowHydrodynamics}.
Here $T=\SI{295}{\kelvin} \pm \SI{2}{\kelvin}$ is the solvent temperature,  $\eta=\SI{0.9}{\milli\pascal\second}$ the dynamic viscosity, $R= \SI{0.77}{\micro\meter} \pm \SI{0.04}{\micro\meter}$ the particle radius, and $\kB$ the Boltzmann constant.
The particle's in-plane orientational ACF is intrinsically multi-exponential~\cite{
  Selmke2018TheoryTransport} and affected by the Janus particle's bottom-heavyness, causing preferential vertical alignment~\cite{OReilly1970RotDiffPolar,Kalmykov1992RotDiffExtPot}. We therefore extract the longest decay time $\tau$ of the orientational ACF by fitting the late-time ACFs with a single exponential function, to get an estimate for $D_r$. In the passive region, we obtained $\tau=\tau_{\te p} = \SI{1.08}{\second} \pm \SI{0.05}{\second}$ [Fig. \ref{fig:figure1}(b)], which is indeed close to the rotational correlation 
  time $1/(2D_{r}) = \SI{1.27}{\second}$, expected for a freely rotating sphere
according to the (rotational) Einstein relation \(
D_{r}
=
\kBT (8\pi\eta R^3)^{-1}
=  \SI{0.39}{\second^{-1}}
\pm \SI{0.06}{\second^{-1}}\)~\cite{debye29PolarMolec}.
Restricting the analysis of the ACF to data points where the particle exhibits strong in-plane alignment, we find a somewhat longer relaxation time $\tau = \SI{2.33}{\second^{-1}} \pm \SI{0.04}{\second^{-1}}$, again in agreement with the expectation for $1/D_{r}$~\cite{
  Selmke2018TheoryTransport}. In the active region, $\tau=\tau_{\te a}$ increases approximately linearly with the laser intensity from $\tau_{\te a}=\SI{1.4}{\second}$ to $\tau_{\te a}=\SI{1.9}{\second}$, which can mainly be attributed to the particle's tendency to increasingly orient in-plane in response to the thermoosmotic flow fields generated by the heating in the confined geometry, and the torques exerted by the radiation pressure~\cite{Das2015BoundariesSpheres,Simmchen2016TopographicalMicroswimmers}.
Finally, the particle's in-plane speed $v_0$ of active self-propulsion along its symmetry axis is deduced from the apparent (long-time) diffusivity  \cite{Cates2015MIPS,Cates2013WhenSeparation}
\begin{equation}\label{eq:deff}
\Deff(x)
= 
D + \frac{v(x)^2}{2\DR(x)} 
\stackrel{x \in \text{bulk}}
{\longrightarrow} 
D + \frac{v_0^2\tau_\te{a}}2,
\end{equation}
in the active region. Both are found to grow nonlinearly with increasing laser intensity, as depicted in Fig.~\ref{fig:figure1}. 
\begin{figure}[tb!]
    \centering
    \includegraphics[width=\columnwidth]{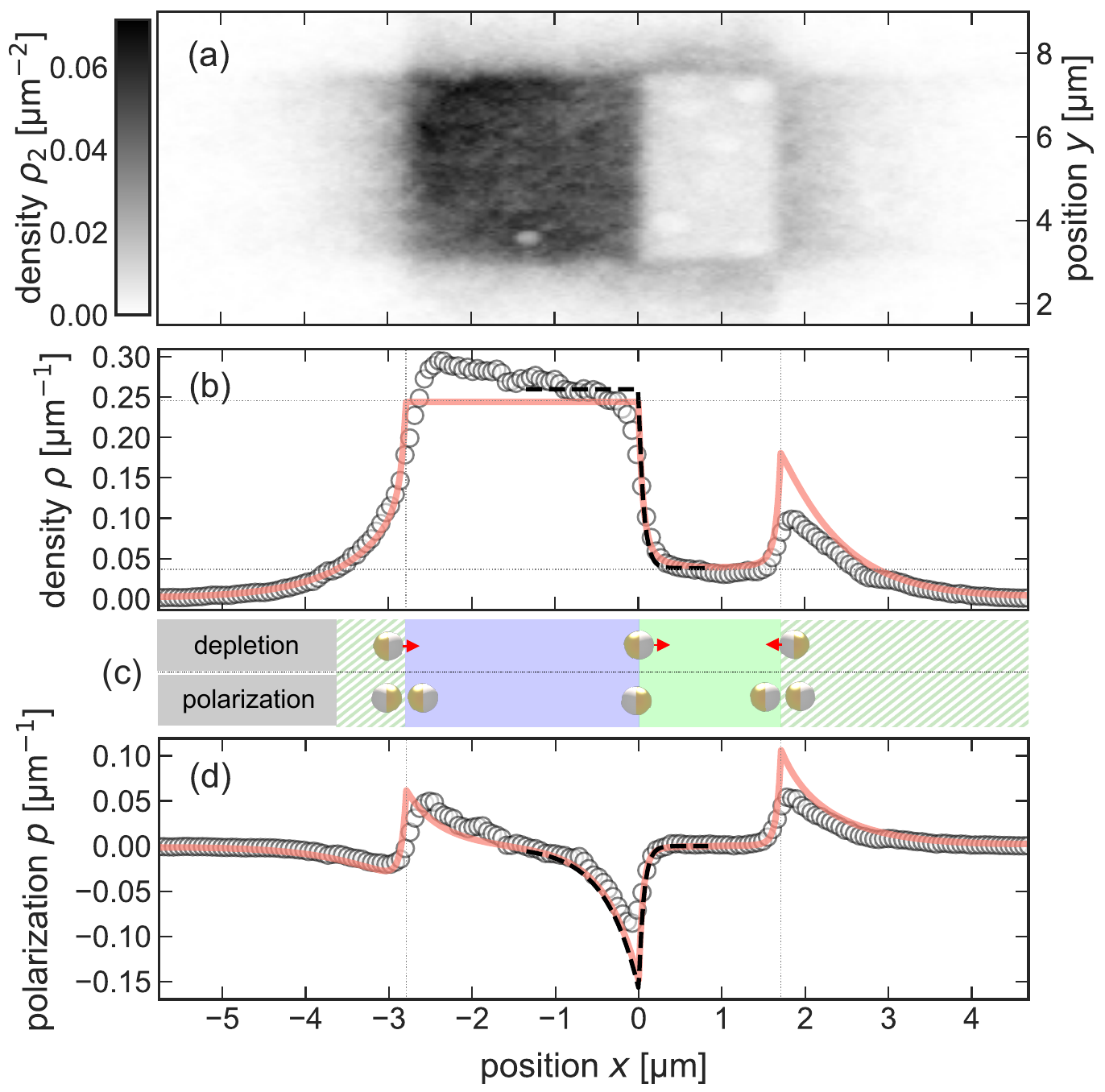}
    \caption{Particle density and polarization.  
    (a) Particle density in the sample plane for $v_0=\SI{2.18}{\micro\m\per\second}$ and $\tau_{\te a}=\SI{1.79}{\second}$ with the active-passive interface at $x=0$.
    (b) Particle density integrated along $y$-direction (excluding a \SI{0.7}{\micro\m} lateral boundary region). The bulk densities $\Rp$, $\Ra$ (upper and lower horizontal line) in the passive/active region were determined at about one polarization decay length, $\lp$, $\la$, respectively, from the passive-active interface. The numerical solution \cite{holubec2019FPESolver} of Eq.~\eqref{eq:fpe} (solid line) is computed using reflecting boundaries. The (approximate) analytic solution from Eqs.~\eqref{eq:ode_rho}-\eqref{eq:rho_ratio} (dashed line) was normalized over half of the nominal width of the active and passive regions. (c) Sketch of the processes creating the interfacial polarization and depletion layers in a simplified  model with binary particle orientations. (d) The experimental, numerical, and analytical particle polarization. The thin dotted vertical lines in (b) and (d) mark the borders to the confining photon-nudging region [cf.~Fig.~\ref{fig:figure1}(a)]).
    }
    \label{fig:figure2}
\end{figure}
\emph{Results:} 
Due to the spatially heterogeneous laser heating in the sample plane, the swim speed of the particle as well as the probability density $\rho_2(x,y)$ to find it at a position $(x,y)$ are spatially heterogeneous. The particle spends 
considerably more time in the passive than in the active region
[Fig.~\ref{fig:figure2}(a)]. Outside of the central arena, the density decays due to photon nudging.
Small imperfections in the data can most likely be attributed to mild statistical fluctuations due to the limited measurement time and localized defects in the Pluronic F127 coating of the cover slips. 
Integrating the density $\rho_2(x,y)$ along the $y$-direction, we obtain the marginal density $\rho(x)$ shown in Fig.~\ref{fig:figure2}(b). It exhibits a pronounced step between essentially homogeneous active and passive bulk plateaus of height $\rho_\te{a,p}$. Further inspection reveals a co-localized excess of Janus orientations $\bf n$ at the density step, pointing along $-{\vv e_{x}}$, towards the passive region (Fig.~\ref{fig:figure2}d). This amounts to a negative average particle polarization
\begin{equation}
  \label{eq:p_def}
  p(x)
  =
  \langle
  {\vv n} \cdot {\vv e_x}
  \rangle
  \rho(x),
\end{equation}
where $\langle . \rangle$ denotes the time average. It decays approximately exponentially with the distance from the activity step. The characteristic decay length on the passive side is substantially longer than that on the active side, $\lp \gg \la$. Similar polarization peaks are seen to occur at the interfaces to the photon-nudging boundary regions, which are regions of a different (orientation-dependent) type of activity~\cite{Auschra2020ActivityFieldsLong}.
Figure~\ref{fig:34_combined}(a) shows that the ratio $\rho_\te{p}/\rho_\te{a}$ of the passive and active bulk density plateaus increases as a function of the experimental estimate $v_0^2 \tau_{\te a}/(2D)$ for the P\'{e}clet number, which characterizes the activity of the Janus particle in the bulk region.
Also the decay lengths $\la$ and $\lp$ depend on the activity contrast --- the former decreasing and the latter slightly increasing (presumably due to the transient wall alignment, thus lower $\DR$, of trespassers) with $v_0$, as seen in Fig.~\ref{fig:34_combined}(b). In summary, the main conclusion drawn from our experiments is that abrupt activity steps are accompanied by (\textit{i}) smooth but pronounced steps in the particle density and (\textit{ii}) the formation of skewed interfacial polarization layers of distinct widths and height.
\begin{figure}[tb!]
    \centering
    \includegraphics[width=\columnwidth]{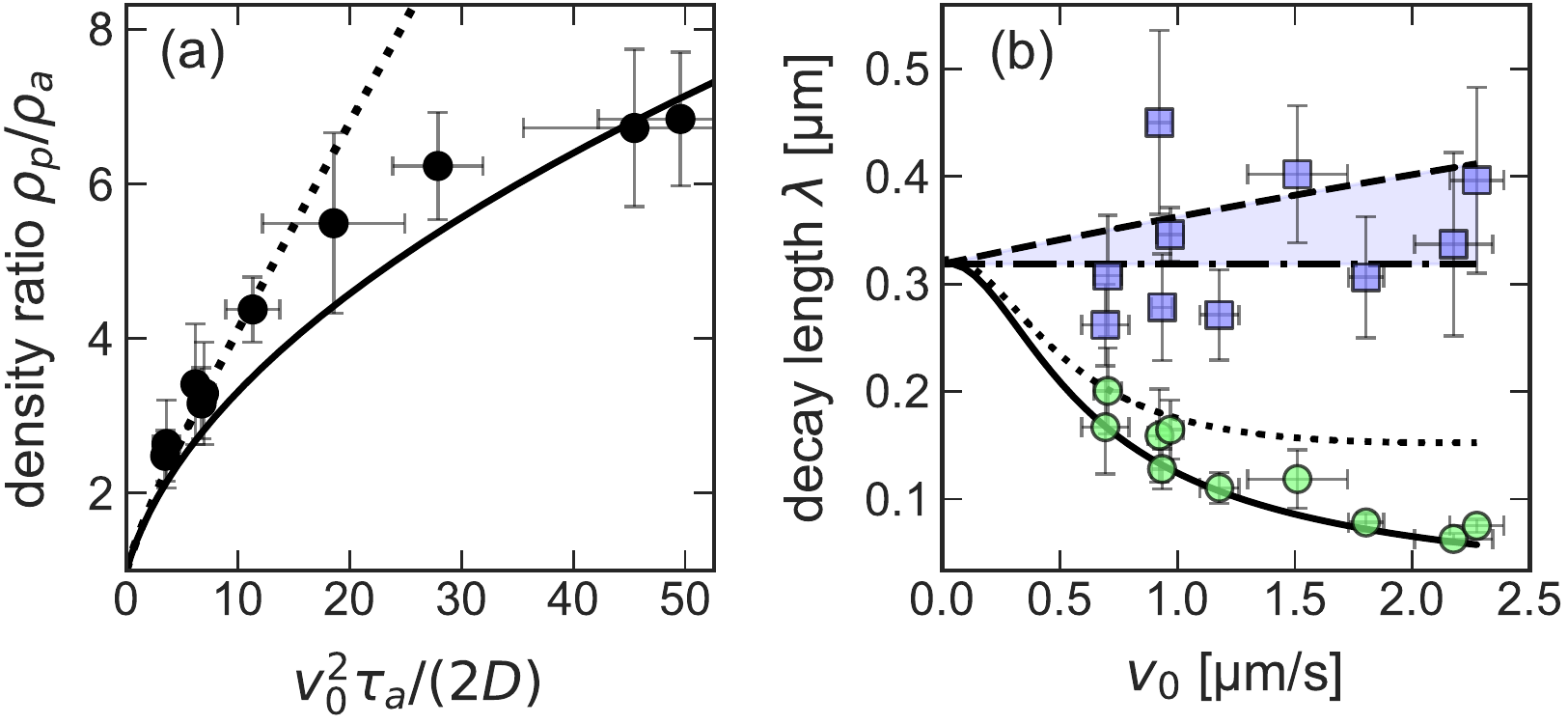}
    \caption{%
    Bulk density ratio and interface widths. (a) Measured ratio $\rho_{p}/\rho_{a}$ of passive and active bulk densities as function of the (experimental) P\'{e}clet number $v_0^2\tau_\te a/2 D$ (circles).  Our analytical prediction~\eqref{eq:rho_ratio} with $\DR=1/\tau_\te{a}$ (solid line) improves that of Ref.~\cite{Fischer2020Quorum-sensingMotility,fischer2020erratum}, namely $\rho_\text{a}/\rho_\text{p}=D_\text{p}/D_{\text a}$ (dotted line).     
      (b) Decay lengths $\lambda_\te{a,p}$ in the active (circles) and passive (squares) bulk regions as functions of $v_0$. The solid and dashed lines show the theoretical prediction~\eqref{eq:lambda} with $\DR=1/\tau_\te{a}$, while the dot-dashed line assumes $\DR=1/\tau_\te{p}$, with $\tau_\te{a,p}$ from the fits in Fig.~\ref{fig:figure1}(b).
      The solid line for $\la$ improves a prediction of Ref.~\cite{Fischer2020Quorum-sensingMotility,fischer2020erratum} (dotted line).
      Error bars indicate the 95\% confidence intervals of $\rho$ and the Gaussian propagation of uncertainties for  $v_0$, $\tau_{\te a}$, $D$ from Fig.~\ref{fig:figure1}, respectively.
    }
    \label{fig:34_combined}
\end{figure}

\emph{Theory:} Our findings can be substantiated by a simple active-Brownian-particle model for the dynamic probability density $f(x,\theta,t)$ to find the Janus particle at time $t$ and position $x$ with orientation angle $\theta \equiv \arccos({\vv n} \cdot {\vv e_x})$ relative to the $x$-axis. Its Fokker-Planck equation reads
\begin{equation}
    \label{eq:fpe}
    \dot f
    =
    D\partial_x^2f
    -
    \partial_x
    \left[
      fv(x)\cos\theta
    \right]
    +
    \DR \partial_\theta^2f,
\end{equation}
where $D$ is the translational diffusion coefficient, $v(x)$ the local propulsion speed, and  $1/\DR$ the orientational correlation time. The stationary solution $f(x,\theta)$ to Eq.~\eqref{eq:fpe} can be approximated by truncating the exact moment expansion with respect to the particle orientation \cite{Cates2013WhenSeparation,Auschra2020ActivityFieldsLong}
after the first two terms, yielding two coupled equations 
\begin{align}
    \label{eq:ode_rho}
    \rho'(x)
    &=
    p(x) v(x)/D,
    \\
    \label{eq:ode_p}
    p''(x)
    &=
    p(x) / \lambda^2(x) + \rho(x)v'(x)/(2D),
\end{align}
for the particle density
\(
\rho(x) 
\equiv 
\int_0^{2\pi} \df \theta ~ 
f(x,\theta)
\)
and the polarization
\(
p(x) 
\equiv 
\int_0^{2\pi} \df \theta ~
f(x,\theta) \cos\theta 
\). 
 For the experimentally realized activity step 
  (see Ref.~\cite{Auschra2020ActivityFieldsLong} for a similar analysis at nudging interfaces)
 we find $\lambda = \lambda_{\rm p}\Theta(-x) + \lambda_{\rm a}\Theta(x)$, with the Heaviside step function $\Theta(x)$ and
\begin{equation}
     \label{eq:lambda}
  \lambda_{\rm p}
   =  
    \bigl(\DR/D\bigr)^{-1/2}~,
  \quad
  \lambda_{\rm a}
  =
    \bigl(\DR/D+v_0^2\,/2D^2\bigr)^{-1/2}.
\end{equation}
In the two regions of constant activity, Eqs.~(\ref{eq:ode_rho}) and (\ref{eq:ode_p}) can be solved exactly~\cite{Auschra2020ActivityFieldsLong}. With the matching condition across the interface, this 
yields the polarization profile  
\begin{equation}
  \label{eq:p_of_x}
  \frac{p_{\te{a,p}}(x)}{\Rp}
  =
  -\frac{v_0}{2D}
  \frac{\la \lp}{\la + \lp}
  \te{e}^{-|x|/\lambda_{\te{a,p}}},
\end{equation}
and (assuming homogeneous $\DR$) the bulk density ratio, 
\begin{equation}
  \label{eq:rho_ratio}  \frac{\rho_\text{a}}{\rho_\text{p}}
  =
  \sqrt{\frac{D_\text{p}}{D_{\text a}}}
  =
  \left(1 + \frac{v_0^2}{2D\DR}\right)^{\!-1/2}  
  =
  \frac{\lambda_\text{a}}{\lambda_\text{p}} \,.
\end{equation} 
Observe that the ratio of the interfacial layer widths $\lambda_\text{p,a}$, the bulk-density ratio $\rho_\text{p}/\rho_\text{a}$, and the reduced peak polarization $p_{\te{a,p}}(0)/\rho_\te{a,p}$ are kinetically determined quantities that only depend on the P\'{e}clet number $v_0^2 /(2D \DR)$. In Ref.~\cite{Auschra2020ActivityFieldsLong}, we detail how the density ratio  can be understood in terms of a detailed balance of two fluxes, a diffusion with the effective diffusivity \eqref{eq:deff} and a nonequilibrium flux due to its spatial heterogeneity.

\emph{Discussion:} A convincing parameter-free comparison of our analytical and experimental findings is obtained if we identify the (homogenous) parameter $\Dr$ of the model with the experimental $1/\tau_\te{a}$. This choice is not entirely trivial, since the theory describes planar rotational motion, whereas the rotational motion of the experimental particle is affected by its mass anisotropy, spatially heterogeneous optical forces, and the excited thermoosmotic flows. Yet, assuming that small statistical uncertainties in position tracking cause some low-pass filtering of the experimental curves, the comparison with the idealized analytical theory in Fig.~\ref{fig:figure2}(b,d)  seems very reasonable.

Intuitively, the (negatively) polarized boundary layer can most easily be understood by a caricature of the above modelling approach in terms of a two-species model that only admits left ($-$) and right ($+$) particle orientations, as sketched in Fig.~\ref{fig:figure2}(c). The net polarization at the activity step is then immediately understood from the quasi-ballistic motion in the active region: it quickly drives the $(+)$ particles to the right edge of the arena and the $(-)$ particles across the interface into the passive region, where they get stuck and cause the negative interface polarization. The active region is thereby depleted relative to the passive region.
The (up to a spurious factor $1/2$) exact analogy between our analytical solutions of Eq.~\eqref{eq:fpe} and the schematic two-species model vindicates our approximation scheme in Eqs.~\eqref{eq:ode_rho}, \eqref{eq:ode_p} \cite{Auschra2020ActivityFieldsLong}. It also explains why, despite the superimposed Brownian motion, Eq.~\eqref{eq:rho_ratio} coincides with the prediction for a run-and-tumble process \cite{Schnitzer1993TheoryChemotaxis}. Our results moreover improve recent theoretical predictions for quorum sensing \cite{Fischer2020Quorum-sensingMotility,fischer2020erratum}. 

Finally, also the skewed shape of the polarization layer is readily explained within the schematic two-species picture. Namely, during the characteristic reorientation time $(2\DR)^{-1}$, an initially perfectly polarized particle starting at the interface diffuses about a distance $\sqrt{2D\tau}=(D/\DR)^{-1/2}$ into the passive region. On the active side, however, the same process is superimposed by self-propulsion, which acts like a sedimentation pressure. It provides a second channel to deplete the negative boundary layer polarization by driving trespassing $(-)$ particles back across the interface. As a result, the interfacial polarization layer on the active side is diminished according to Eq.~\eqref{eq:lambda}, which adds the two decay channels together. 
The total polarization 
\begin{equation}
  \label{eq:Pol_total}
  P_\text{tot}
  =
  \int_{-\infty}^\infty \text{d}x ~ p(x)
  = -
  \frac{v_0 \rho_\text{a}}{2 \DR} < 0 \;,
\end{equation}
is a state function \cite{hermann2020PolStateFct,Auschra2020ActivityFieldsLong}, completely defined by the magnitude of bulk currents (here only $v_0 \rho_\text{a}\neq 0$) and $\DR$, as a glance at Eqs.~\eqref{eq:lambda}-\eqref{eq:rho_ratio} confirms. 
Since $P_\text{tot}$ also determines the ``swim pressure'' exerted by the active particle onto the dense bulk phase~\cite{Solon2015PressureSpheres}, both are entirely determined by the bulk currents.
There does not seem to be any straightforward explanation in terms of a static equilibrium analogy~\cite{solon2018GenTDofMIPS} in the sense of a phase coexistence maintained by swim pressure, though, suggesting that polarization-density patterns can play the role of a smoking gun for particle-level activity.

In summary, we have employed the sophisticated technique of photon nudging to set up a boundary-free activity arena, thereby establishing a potent test bed to address fundamental open issues in  active-particle physics. It enables us to observe active-particle motion in activity landscapes over several days. We found that motility gradients are accompanied by characteristic skewed interfacial polarization profiles. We showed them to arise from active-particle fluxes in the bulk and to constitute a  state  function  of  the  nonequilibrium  stationary  state,  without  admitting any straightforward  analogy with equilibrium phase coexistence \cite{caprini2020velocityAlignment}. Our experiments are well described by a precise quantitative theory that advances previous work, can be generalized to photon-nudging, and can deal with further phenomena such as wall accumulation and quorum-sensing in active phase transitions \cite{Auschra2020ActivityFieldsLong}. It allows essential microscopic parameters such as swim speed and (effective) translational and rotational diffusion coefficients to be inferred from accessible mesoscopic observables, namely the bulk particle-density and interfacial polarization profiles. And it suggests that similar polarization-density patterns are a hallmark of all kinds of hustle and bustle of animalcules in heterogeneous activity landscapes, far beyond our artificial model system.
As our experimental approach is capable of handling a controlled number of of active particles simultaneously, a challenging but interesting avenue for future research would be to try and extend our experiments and theory to interacting many-body problems.

  \section*{Acknowledgements}
\label{sec:acknowledgements}

We acknowledge funding by the Deutsche Forschungsgemeinschaft (DFG) through the priority program ``Microswimmers'' (SPP 1726, project 237143019) and by the Czech Science Foundation (GACR project 20-02955J). V.~H. was supported by the Humboldt foundation and N.~A.~S. by the European Science Foundation (ESF project 100327895).

\bibliography{references} 

\begin{thebibliography}{39}%
\makeatletter
\providecommand \@ifxundefined [1]{%
 \@ifx{#1\undefined}
}%
\providecommand \@ifnum [1]{%
 \ifnum #1\expandafter \@firstoftwo
 \else \expandafter \@secondoftwo
 \fi
}%
\providecommand \@ifx [1]{%
 \ifx #1\expandafter \@firstoftwo
 \else \expandafter \@secondoftwo
 \fi
}%
\providecommand \natexlab [1]{#1}%
\providecommand \enquote  [1]{``#1''}%
\providecommand \bibnamefont  [1]{#1}%
\providecommand \bibfnamefont [1]{#1}%
\providecommand \citenamefont [1]{#1}%
\providecommand \href@noop [0]{\@secondoftwo}%
\providecommand \href [0]{\begingroup \@sanitize@url \@href}%
\providecommand \@href[1]{\@@startlink{#1}\@@href}%
\providecommand \@@href[1]{\endgroup#1\@@endlink}%
\providecommand \@sanitize@url [0]{\catcode `\\12\catcode `\$12\catcode
  `\&12\catcode `\#12\catcode `\^12\catcode `\_12\catcode `\%12\relax}%
\providecommand \@@startlink[1]{}%
\providecommand \@@endlink[0]{}%
\providecommand \url  [0]{\begingroup\@sanitize@url \@url }%
\providecommand \@url [1]{\endgroup\@href {#1}{\urlprefix }}%
\providecommand \urlprefix  [0]{URL }%
\providecommand \Eprint [0]{\href }%
\providecommand \doibase [0]{http://dx.doi.org/}%
\providecommand \selectlanguage [0]{\@gobble}%
\providecommand \bibinfo  [0]{\@secondoftwo}%
\providecommand \bibfield  [0]{\@secondoftwo}%
\providecommand \translation [1]{[#1]}%
\providecommand \BibitemOpen [0]{}%
\providecommand \bibitemStop [0]{}%
\providecommand \bibitemNoStop [0]{.\EOS\space}%
\providecommand \EOS [0]{\spacefactor3000\relax}%
\providecommand \BibitemShut  [1]{\csname bibitem#1\endcsname}%
\let\auto@bib@innerbib\@empty
\bibitem [{\citenamefont {Lane}(2015)}]{Lane2015Animalcules}%
  \BibitemOpen
  \bibfield  {author} {\bibinfo {author} {\bibfnamefont {N.}~\bibnamefont
  {Lane}},\ }\href {https://doi.org/10.1098/rstb.2014.0380} {\bibfield
  {journal} {\bibinfo  {journal} {Philosophical Transactions of the Royal
  Society B: Biological Sciences}\ }\textbf {\bibinfo {volume} {370}},\
  \bibinfo {pages} {20140344} (\bibinfo {year} {2015})}\BibitemShut {NoStop}%
\bibitem [{\citenamefont {Cates}(2012)}]{Cates2012DiffusivePhysics}%
  \BibitemOpen
  \bibfield  {author} {\bibinfo {author} {\bibfnamefont {M.~E.}\ \bibnamefont
  {Cates}},\ }\href {https://doi.org/10.1088/0034-4885/75/4/042601} {\bibfield
  {journal} {\bibinfo  {journal} {Reports on Progress in Physics}\ }\textbf
  {\bibinfo {volume} {75}},\ \bibinfo {pages} {042601} (\bibinfo {year}
  {2012})}\BibitemShut {NoStop}%
\bibitem [{\citenamefont {Contera}(2019)}]{Contera2019Nano}%
  \BibitemOpen
  \bibfield  {author} {\bibinfo {author} {\bibfnamefont {S.}~\bibnamefont
  {Contera}},\ }\href@noop {} {\emph {\bibinfo {title} {Nano comes to life :
  how nanotechnology is transforming medicine and the future of biology}}}\
  (\bibinfo  {publisher} {Princeton University Press},\ \bibinfo {address}
  {Princeton},\ \bibinfo {year} {2019})\BibitemShut {NoStop}%
\bibitem [{\citenamefont {Gnesotto}\ \emph {et~al.}(2018)\citenamefont
  {Gnesotto}, \citenamefont {Mura}, \citenamefont {Gladrow},\ and\
  \citenamefont {Broedersz}}]{Gnesotto2018BrokenDetailedBalance}%
  \BibitemOpen
  \bibfield  {author} {\bibinfo {author} {\bibfnamefont {F.~S.}\ \bibnamefont
  {Gnesotto}}, \bibinfo {author} {\bibfnamefont {F.}~\bibnamefont {Mura}},
  \bibinfo {author} {\bibfnamefont {J.}~\bibnamefont {Gladrow}}, \ and\
  \bibinfo {author} {\bibfnamefont {C.~P.}\ \bibnamefont {Broedersz}},\ }\href
  {\doibase 10.1088/1361-6633/aab3ed} {\bibfield  {journal} {\bibinfo
  {journal} {Reports on Progress in Physics}\ }\textbf {\bibinfo {volume}
  {81}},\ \bibinfo {pages} {066601} (\bibinfo {year} {2018})}\BibitemShut
  {NoStop}%
\bibitem [{\citenamefont {Steffenoni}\ \emph {et~al.}(2016)\citenamefont
  {Steffenoni}, \citenamefont {Kroy},\ and\ \citenamefont
  {Falasco}}]{Steffenoni2016InteractingBath}%
  \BibitemOpen
  \bibfield  {author} {\bibinfo {author} {\bibfnamefont {S.}~\bibnamefont
  {Steffenoni}}, \bibinfo {author} {\bibfnamefont {K.}~\bibnamefont {Kroy}}, \
  and\ \bibinfo {author} {\bibfnamefont {G.}~\bibnamefont {Falasco}},\ }\href
  {https://doi.org/10.1103/PhysRevE.94.062139} {\bibfield  {journal} {\bibinfo
  {journal} {Physical Review E}\ }\textbf {\bibinfo {volume} {94}},\ \bibinfo
  {pages} {062139} (\bibinfo {year} {2016})}\BibitemShut {NoStop}%
\bibitem [{\citenamefont {Basu}\ \emph {et~al.}(2015)\citenamefont {Basu},
  \citenamefont {Maes},\ and\ \citenamefont
  {Neto\v{c}n\'{y}}}]{Basu2015StatForces}%
  \BibitemOpen
  \bibfield  {author} {\bibinfo {author} {\bibfnamefont {U.}~\bibnamefont
  {Basu}}, \bibinfo {author} {\bibfnamefont {C.}~\bibnamefont {Maes}}, \ and\
  \bibinfo {author} {\bibfnamefont {K.}~\bibnamefont {Neto\v{c}n\'{y}}},\
  }\href {https://doi.org/10.1103/PhysRevLett.114.250601} {\bibfield  {journal}
  {\bibinfo  {journal} {Physical Review Letters}\ }\textbf {\bibinfo {volume}
  {114}},\ \bibinfo {pages} {250601} (\bibinfo {year} {2015})}\BibitemShut
  {NoStop}%
\bibitem [{\citenamefont {Solon}\ \emph
  {et~al.}(2018{\natexlab{a}})\citenamefont {Solon}, \citenamefont
  {Stenhammar}, \citenamefont {Cates}, \citenamefont {Kafri},\ and\
  \citenamefont {Tailleur}}]{Solon2018GeneralizedMatter}%
  \BibitemOpen
  \bibfield  {author} {\bibinfo {author} {\bibfnamefont {A.~P.}\ \bibnamefont
  {Solon}}, \bibinfo {author} {\bibfnamefont {J.}~\bibnamefont {Stenhammar}},
  \bibinfo {author} {\bibfnamefont {M.~E.}\ \bibnamefont {Cates}}, \bibinfo
  {author} {\bibfnamefont {Y.}~\bibnamefont {Kafri}}, \ and\ \bibinfo {author}
  {\bibfnamefont {J.}~\bibnamefont {Tailleur}},\ }\href {\doibase
  10.1103/PhysRevE.97.020602} {\bibfield  {journal} {\bibinfo  {journal}
  {Physical Review E}\ }\textbf {\bibinfo {volume} {97}},\ \bibinfo {pages}
  {020602(R)} (\bibinfo {year} {2018}{\natexlab{a}})}\BibitemShut {NoStop}%
\bibitem [{\citenamefont {Cates}\ and\ \citenamefont
  {Tailleur}(2015)}]{Cates2015MIPS}%
  \BibitemOpen
  \bibfield  {author} {\bibinfo {author} {\bibfnamefont {M.~E.}\ \bibnamefont
  {Cates}}\ and\ \bibinfo {author} {\bibfnamefont {J.}~\bibnamefont
  {Tailleur}},\ }\href {\doibase 10.1146/annurev-conmatphys-031214-014710}
  {\bibfield  {journal} {\bibinfo  {journal} {Annual Review of Condensed Matter
  Physics}\ }\textbf {\bibinfo {volume} {6}},\ \bibinfo {pages} {219–244}
  (\bibinfo {year} {2015})}\BibitemShut {NoStop}%
\bibitem [{\citenamefont {Cates}\ and\ \citenamefont
  {Tailleur}(2013)}]{Cates2013WhenSeparation}%
  \BibitemOpen
  \bibfield  {author} {\bibinfo {author} {\bibfnamefont {M.~E.}\ \bibnamefont
  {Cates}}\ and\ \bibinfo {author} {\bibfnamefont {J.}~\bibnamefont
  {Tailleur}},\ }\href {https://doi.org/10.1209/0295-5075/101/20010} {\bibfield
   {journal} {\bibinfo  {journal} {EPL}\ }\textbf {\bibinfo {volume} {101}},\
  \bibinfo {pages} {20010} (\bibinfo {year} {2013})}\BibitemShut {NoStop}%
\bibitem [{\citenamefont {Shida}\ and\ \citenamefont
  {Kawai}(1989)}]{Shida1989InelasticCollisions}%
  \BibitemOpen
  \bibfield  {author} {\bibinfo {author} {\bibfnamefont {K.}~\bibnamefont
  {Shida}}\ and\ \bibinfo {author} {\bibfnamefont {T.}~\bibnamefont {Kawai}},\
  }\href {\doibase 10.1016/0378-4371(89)90562-1} {\bibfield  {journal}
  {\bibinfo  {journal} {Physica A: Statistical Mechanics and its Applications}\
  }\textbf {\bibinfo {volume} {162}},\ \bibinfo {pages} {145–160} (\bibinfo
  {year} {1989})}\BibitemShut {NoStop}%
\bibitem [{\citenamefont {McNamara}\ and\ \citenamefont
  {Young}(1992)}]{McNamara1992InelasticCollapse}%
  \BibitemOpen
  \bibfield  {author} {\bibinfo {author} {\bibfnamefont {S.}~\bibnamefont
  {McNamara}}\ and\ \bibinfo {author} {\bibfnamefont {W.~R.}\ \bibnamefont
  {Young}},\ }\href {\doibase 10.1063/1.858323} {\bibfield  {journal} {\bibinfo
   {journal} {Physics of Fluids A: Fluid Dynamics}\ }\textbf {\bibinfo {volume}
  {4}},\ \bibinfo {pages} {496–504} (\bibinfo {year} {1992})}\BibitemShut
  {NoStop}%
\bibitem [{\citenamefont {Schnitzer}(1993)}]{Schnitzer1993TheoryChemotaxis}%
  \BibitemOpen
  \bibfield  {author} {\bibinfo {author} {\bibfnamefont {M.~J.}\ \bibnamefont
  {Schnitzer}},\ }\href {\doibase 10.1103/PhysRevE.48.2553} {\bibfield
  {journal} {\bibinfo  {journal} {Physical Review E}\ }\textbf {\bibinfo
  {volume} {48}},\ \bibinfo {pages} {2553} (\bibinfo {year}
  {1993})}\BibitemShut {NoStop}%
\bibitem [{\citenamefont {Weber}\ \emph {et~al.}(2016)\citenamefont {Weber},
  \citenamefont {Weber},\ and\ \citenamefont
  {Frey}}]{Weber2016BinaryMixturesDemix}%
  \BibitemOpen
  \bibfield  {author} {\bibinfo {author} {\bibfnamefont {S.~N.}\ \bibnamefont
  {Weber}}, \bibinfo {author} {\bibfnamefont {C.~A.}\ \bibnamefont {Weber}}, \
  and\ \bibinfo {author} {\bibfnamefont {E.}~\bibnamefont {Frey}},\ }\href
  {https://doi.org/10.1103/PhysRevLett.116.058301} {\bibfield  {journal}
  {\bibinfo  {journal} {Physical Review Letters}\ }\textbf {\bibinfo {volume}
  {116}},\ \bibinfo {pages} {058301} (\bibinfo {year} {2016})}\BibitemShut
  {NoStop}%
\bibitem [{\citenamefont {Elgeti}\ and\ \citenamefont
  {Gompper}(2013)}]{Elgeti2013WallAccumulation}%
  \BibitemOpen
  \bibfield  {author} {\bibinfo {author} {\bibfnamefont {J.}~\bibnamefont
  {Elgeti}}\ and\ \bibinfo {author} {\bibfnamefont {G.}~\bibnamefont
  {Gompper}},\ }\href {\doibase 10.1209/0295-5075/101/48003} {\bibfield
  {journal} {\bibinfo  {journal} {EPL}\ }\textbf {\bibinfo {volume} {101}},\
  \bibinfo {pages} {48003} (\bibinfo {year} {2013})}\BibitemShut {NoStop}%
\bibitem [{\citenamefont {Speck}\ and\ \citenamefont
  {Jack}(2016)}]{Speck2016IdealBulkPressure}%
  \BibitemOpen
  \bibfield  {author} {\bibinfo {author} {\bibfnamefont {T.}~\bibnamefont
  {Speck}}\ and\ \bibinfo {author} {\bibfnamefont {R.~L.}\ \bibnamefont
  {Jack}},\ }\href {https://doi.org/10.1103/PhysRevE.93.062605} {\bibfield
  {journal} {\bibinfo  {journal} {Physical Review E}\ }\textbf {\bibinfo
  {volume} {93}},\ \bibinfo {pages} {062605} (\bibinfo {year}
  {2016})}\BibitemShut {NoStop}%
\bibitem [{\citenamefont {Wagner}\ \emph {et~al.}(2017)\citenamefont {Wagner},
  \citenamefont {Hagan},\ and\ \citenamefont
  {Baskaran}}]{Wagner2017ABPsUnderConfinement}%
  \BibitemOpen
  \bibfield  {author} {\bibinfo {author} {\bibfnamefont {C.~G.}\ \bibnamefont
  {Wagner}}, \bibinfo {author} {\bibfnamefont {M.~F.}\ \bibnamefont {Hagan}}, \
  and\ \bibinfo {author} {\bibfnamefont {A.}~\bibnamefont {Baskaran}},\ }\href
  {\doibase 10.1088/1742-5468/aa60a8} {\bibfield  {journal} {\bibinfo
  {journal} {Journal of Statistical Mechanics: Theory and Experiment}\ }\textbf
  {\bibinfo {volume} {2017}},\ \bibinfo {pages} {043203} (\bibinfo {year}
  {2017})}\BibitemShut {NoStop}%
\bibitem [{\citenamefont {Seifert}(2012)}]{Seifert2012MolecularMachines}%
  \BibitemOpen
  \bibfield  {author} {\bibinfo {author} {\bibfnamefont {U.}~\bibnamefont
  {Seifert}},\ }\href {\doibase 10.1088/0034-4885/75/12/126001} {\bibfield
  {journal} {\bibinfo  {journal} {Reports on Progress in Physics}\ }\textbf
  {\bibinfo {volume} {75}},\ \bibinfo {pages} {126001} (\bibinfo {year}
  {2012})}\BibitemShut {NoStop}%
\bibitem [{\citenamefont {Holubec}\ \emph {et~al.}(2017)\citenamefont
  {Holubec}, \citenamefont {Ryabov}, \citenamefont {Yaghoubi}, \citenamefont
  {Varga}, \citenamefont {Khodaee}, \citenamefont {Foulaadvand},\ and\
  \citenamefont {Chvosta}}]{Holubec2017ThermalRatchetConfGeom}%
  \BibitemOpen
  \bibfield  {author} {\bibinfo {author} {\bibfnamefont {V.}~\bibnamefont
  {Holubec}}, \bibinfo {author} {\bibfnamefont {A.}~\bibnamefont {Ryabov}},
  \bibinfo {author} {\bibfnamefont {M.}~\bibnamefont {Yaghoubi}}, \bibinfo
  {author} {\bibfnamefont {M.}~\bibnamefont {Varga}}, \bibinfo {author}
  {\bibfnamefont {A.}~\bibnamefont {Khodaee}}, \bibinfo {author} {\bibfnamefont
  {M.}~\bibnamefont {Foulaadvand}}, \ and\ \bibinfo {author} {\bibfnamefont
  {P.}~\bibnamefont {Chvosta}},\ }\href {\doibase 10.3390/e19040119} {\bibfield
   {journal} {\bibinfo  {journal} {Entropy}\ }\textbf {\bibinfo {volume}
  {19}},\ \bibinfo {pages} {119} (\bibinfo {year} {2017})}\BibitemShut
  {NoStop}%
\bibitem [{\citenamefont {Lieleg}\ \emph {et~al.}(2007)\citenamefont {Lieleg},
  \citenamefont {Claessens}, \citenamefont {Heussinger}, \citenamefont {Frey},\
  and\ \citenamefont {Bausch}}]{Lieleg2007BundlesSemiFlexPolymers}%
  \BibitemOpen
  \bibfield  {author} {\bibinfo {author} {\bibfnamefont {O.}~\bibnamefont
  {Lieleg}}, \bibinfo {author} {\bibfnamefont {M.~M. A.~E.}\ \bibnamefont
  {Claessens}}, \bibinfo {author} {\bibfnamefont {C.}~\bibnamefont
  {Heussinger}}, \bibinfo {author} {\bibfnamefont {E.}~\bibnamefont {Frey}}, \
  and\ \bibinfo {author} {\bibfnamefont {A.~R.}\ \bibnamefont {Bausch}},\
  }\href {https://doi.org/10.1103/PhysRevLett.99.088102} {\bibfield  {journal}
  {\bibinfo  {journal} {Physical Review Letters}\ }\textbf {\bibinfo {volume}
  {99}},\ \bibinfo {pages} {088102} (\bibinfo {year} {2007})}\BibitemShut
  {NoStop}%
\bibitem [{\citenamefont {Auschra}\ \emph {et~al.}(2020)\citenamefont
  {Auschra}, \citenamefont {S{\"o}ker}, \citenamefont {Holubec}, \citenamefont
  {Cichos},\ and\ \citenamefont {Kroy}}]{Auschra2020ActivityFieldsLong}%
  \BibitemOpen
  \bibfield  {author} {\bibinfo {author} {\bibfnamefont {S.}~\bibnamefont
  {Auschra}}, \bibinfo {author} {\bibfnamefont {N.}~\bibnamefont {S{\"o}ker}},
  \bibinfo {author} {\bibfnamefont {V.}~\bibnamefont {Holubec}}, \bibinfo
  {author} {\bibfnamefont {F.}~\bibnamefont {Cichos}}, \ and\ \bibinfo {author}
  {\bibfnamefont {K.}~\bibnamefont {Kroy}},\ }\href@noop {} {\enquote {\bibinfo
  {title} {Polariazation-density patterns of active particles in motility
  gradients},}\ } (\bibinfo {year} {2020}),\ \bibinfo {note} {companion article
  submitted to Phys. Rev. E}\BibitemShut {NoStop}%
\bibitem [{\citenamefont {Qian}\ \emph {et~al.}(2013)\citenamefont {Qian},
  \citenamefont {Montiel}, \citenamefont {Bregulla}, \citenamefont {Cichos},\
  and\ \citenamefont {Yang}}]{Qian2013HarnessingNudgingb}%
  \BibitemOpen
  \bibfield  {author} {\bibinfo {author} {\bibfnamefont {B.}~\bibnamefont
  {Qian}}, \bibinfo {author} {\bibfnamefont {D.}~\bibnamefont {Montiel}},
  \bibinfo {author} {\bibfnamefont {A.}~\bibnamefont {Bregulla}}, \bibinfo
  {author} {\bibfnamefont {F.}~\bibnamefont {Cichos}}, \ and\ \bibinfo {author}
  {\bibfnamefont {H.}~\bibnamefont {Yang}},\ }\href {\doibase
  10.1039/c2sc21263c} {\bibfield  {journal} {\bibinfo  {journal} {Chemical
  Science}\ }\textbf {\bibinfo {volume} {4}},\ \bibinfo {pages} {1420}
  (\bibinfo {year} {2013})}\BibitemShut {NoStop}%
\bibitem [{\citenamefont {Bregulla}\ \emph {et~al.}(2014)\citenamefont
  {Bregulla}, \citenamefont {Yang},\ and\ \citenamefont
  {Cichos}}]{Bregulla2014StochasticNudging}%
  \BibitemOpen
  \bibfield  {author} {\bibinfo {author} {\bibfnamefont {A.~P.}\ \bibnamefont
  {Bregulla}}, \bibinfo {author} {\bibfnamefont {H.}~\bibnamefont {Yang}}, \
  and\ \bibinfo {author} {\bibfnamefont {F.}~\bibnamefont {Cichos}},\ }\href
  {https://doi.org/10.1021/nn501568e} {\bibfield  {journal} {\bibinfo
  {journal} {ACS Nano}\ }\textbf {\bibinfo {volume} {8}},\ \bibinfo {pages}
  {6542} (\bibinfo {year} {2014})}\BibitemShut {NoStop}%
\bibitem [{\citenamefont {Selmke}\ \emph
  {et~al.}(2018{\natexlab{a}})\citenamefont {Selmke}, \citenamefont {Khadka},
  \citenamefont {Bregulla}, \citenamefont {Cichos},\ and\ \citenamefont
  {Yang}}]{Selmke2018TheoryTransport}%
  \BibitemOpen
  \bibfield  {author} {\bibinfo {author} {\bibfnamefont {M.}~\bibnamefont
  {Selmke}}, \bibinfo {author} {\bibfnamefont {U.}~\bibnamefont {Khadka}},
  \bibinfo {author} {\bibfnamefont {A.~P.}\ \bibnamefont {Bregulla}}, \bibinfo
  {author} {\bibfnamefont {F.}~\bibnamefont {Cichos}}, \ and\ \bibinfo {author}
  {\bibfnamefont {H.}~\bibnamefont {Yang}},\ }\href {\doibase
  10.1039/c7cp06559k} {\bibfield  {journal} {\bibinfo  {journal} {Physical
  Chemistry Chemical Physics}\ }\textbf {\bibinfo {volume} {20}},\ \bibinfo
  {pages} {10502} (\bibinfo {year} {2018}{\natexlab{a}})}\BibitemShut {NoStop}%
\bibitem [{\citenamefont {Selmke}\ \emph
  {et~al.}(2018{\natexlab{b}})\citenamefont {Selmke}, \citenamefont {Khadka},
  \citenamefont {Bregulla}, \citenamefont {Cichos},\ and\ \citenamefont
  {Yang}}]{Selmke2018TheoryConfinement}%
  \BibitemOpen
  \bibfield  {author} {\bibinfo {author} {\bibfnamefont {M.}~\bibnamefont
  {Selmke}}, \bibinfo {author} {\bibfnamefont {U.}~\bibnamefont {Khadka}},
  \bibinfo {author} {\bibfnamefont {A.~P.}\ \bibnamefont {Bregulla}}, \bibinfo
  {author} {\bibfnamefont {F.}~\bibnamefont {Cichos}}, \ and\ \bibinfo {author}
  {\bibfnamefont {H.}~\bibnamefont {Yang}},\ }\href {\doibase
  10.1039/c7cp06560d} {\bibfield  {journal} {\bibinfo  {journal} {Physical
  Chemistry Chemical Physics}\ }\textbf {\bibinfo {volume} {20}},\ \bibinfo
  {pages} {10521} (\bibinfo {year} {2018}{\natexlab{b}})}\BibitemShut {NoStop}%
\bibitem [{\citenamefont {Hermann}\ and\ \citenamefont
  {Schmidt}(2020)}]{hermann2020PolStateFct}%
  \BibitemOpen
  \bibfield  {author} {\bibinfo {author} {\bibfnamefont {S.}~\bibnamefont
  {Hermann}}\ and\ \bibinfo {author} {\bibfnamefont {M.}~\bibnamefont
  {Schmidt}},\ }\href {https://doi.org/10.1103/PhysRevResearch.2.022003}
  {\bibfield  {journal} {\bibinfo  {journal} {Physical Review Research}\
  }\textbf {\bibinfo {volume} {2}},\ \bibinfo {pages} {022003(R)} (\bibinfo
  {year} {2020})}\BibitemShut {NoStop}%
\bibitem [{\citenamefont {Malakar}\ \emph {et~al.}(2018)\citenamefont
  {Malakar}, \citenamefont {Jemseena}, \citenamefont {Kundu}, \citenamefont
  {Vijay~Kumar}, \citenamefont {Sabhapandit}, \citenamefont {Majumdar},
  \citenamefont {Redner},\ and\ \citenamefont
  {Dhar}}]{Malakar2018RunAndTumble1D}%
  \BibitemOpen
  \bibfield  {author} {\bibinfo {author} {\bibfnamefont {K.}~\bibnamefont
  {Malakar}}, \bibinfo {author} {\bibfnamefont {V.}~\bibnamefont {Jemseena}},
  \bibinfo {author} {\bibfnamefont {A.}~\bibnamefont {Kundu}}, \bibinfo
  {author} {\bibfnamefont {K.}~\bibnamefont {Vijay~Kumar}}, \bibinfo {author}
  {\bibfnamefont {S.}~\bibnamefont {Sabhapandit}}, \bibinfo {author}
  {\bibfnamefont {S.~N.}\ \bibnamefont {Majumdar}}, \bibinfo {author}
  {\bibfnamefont {S.}~\bibnamefont {Redner}}, \ and\ \bibinfo {author}
  {\bibfnamefont {A.}~\bibnamefont {Dhar}},\ }\href {\doibase
  10.1088/1742-5468/aab84f} {\bibfield  {journal} {\bibinfo  {journal} {Journal
  of Statistical Mechanics: Theory and Experiment}\ }\textbf {\bibinfo {volume}
  {2018}},\ \bibinfo {pages} {043215} (\bibinfo {year} {2018})}\BibitemShut
  {NoStop}%
\bibitem [{\citenamefont {Sharma}\ and\ \citenamefont
  {Brader}(2017)}]{Sharma2017BrownianActivity}%
  \BibitemOpen
  \bibfield  {author} {\bibinfo {author} {\bibfnamefont {A.}~\bibnamefont
  {Sharma}}\ and\ \bibinfo {author} {\bibfnamefont {J.~M.}\ \bibnamefont
  {Brader}},\ }\href {https://doi.org/10.1103/PhysRevE.96.032604} {\bibfield
  {journal} {\bibinfo  {journal} {Physical Review E}\ }\textbf {\bibinfo
  {volume} {96}},\ \bibinfo {pages} {032604} (\bibinfo {year}
  {2017})}\BibitemShut {NoStop}%
\bibitem [{\citenamefont {Fischer}\ \emph
  {et~al.}(2020{\natexlab{a}})\citenamefont {Fischer}, \citenamefont {Schmid},\
  and\ \citenamefont {Speck}}]{Fischer2020Quorum-sensingMotility}%
  \BibitemOpen
  \bibfield  {author} {\bibinfo {author} {\bibfnamefont {A.}~\bibnamefont
  {Fischer}}, \bibinfo {author} {\bibfnamefont {F.}~\bibnamefont {Schmid}}, \
  and\ \bibinfo {author} {\bibfnamefont {T.}~\bibnamefont {Speck}},\ }\href
  {https://doi.org/10.1103/PhysRevE.101.012601} {\bibfield  {journal} {\bibinfo
   {journal} {Physical Review E}\ }\textbf {\bibinfo {volume} {101}},\ \bibinfo
  {pages} {012601} (\bibinfo {year} {2020}{\natexlab{a}})}\BibitemShut
  {NoStop}%
\bibitem [{\citenamefont {Fischer}\ \emph
  {et~al.}(2020{\natexlab{b}})\citenamefont {Fischer}, \citenamefont {Schmid},\
  and\ \citenamefont {Speck}}]{fischer2020erratum}%
  \BibitemOpen
  \bibfield  {author} {\bibinfo {author} {\bibfnamefont {A.}~\bibnamefont
  {Fischer}}, \bibinfo {author} {\bibfnamefont {F.}~\bibnamefont {Schmid}}, \
  and\ \bibinfo {author} {\bibfnamefont {T.}~\bibnamefont {Speck}},\ }\href
  {http://dx.doi.org/10.1103/PhysRevE.102.059903} {\bibfield  {journal}
  {\bibinfo  {journal} {Physical Review E}\ }\textbf {\bibinfo {volume}
  {102}},\ \bibinfo {pages} {059903(E)} (\bibinfo {year}
  {2020}{\natexlab{b}})}\BibitemShut {NoStop}%
\bibitem [{\citenamefont {Happel}\ and\ \citenamefont
  {Brenner}(1981)}]{Happel1981LowHydrodynamics}%
  \BibitemOpen
  \bibfield  {author} {\bibinfo {author} {\bibfnamefont {J.}~\bibnamefont
  {Happel}}\ and\ \bibinfo {author} {\bibfnamefont {H.}~\bibnamefont
  {Brenner}},\ }\href
  {https://www.cambridge.org/core/product/identifier/CBO9781316134030A075/type/book_part
  http://link.springer.com/10.1007/978-94-009-8352-6} {\emph {\bibinfo {title}
  {Low Reynolds number hydrodynamics}}},\ \bibinfo {series} {Mechanics of
  fluids and transport processes}, Vol.~\bibinfo {volume} {1}\ (\bibinfo
  {publisher} {Springer Netherlands},\ \bibinfo {address} {Dordrecht},\
  \bibinfo {year} {1981})\BibitemShut {NoStop}%
\bibitem [{\citenamefont {O’Reilly}(1970)}]{OReilly1970RotDiffPolar}%
  \BibitemOpen
  \bibfield  {author} {\bibinfo {author} {\bibfnamefont {D.~E.}\ \bibnamefont
  {O’Reilly}},\ }\href {\doibase 10.1021/j100711a021} {\bibfield  {journal}
  {\bibinfo  {journal} {The Journal of Physical Chemistry}\ }\textbf {\bibinfo
  {volume} {74}},\ \bibinfo {pages} {3277–3279} (\bibinfo {year}
  {1970})}\BibitemShut {NoStop}%
\bibitem [{\citenamefont {Kalmykov}(1992)}]{Kalmykov1992RotDiffExtPot}%
  \BibitemOpen
  \bibfield  {author} {\bibinfo {author} {\bibfnamefont {Y.~P.}\ \bibnamefont
  {Kalmykov}},\ }\href {\doibase 10.1103/physreva.45.7184} {\bibfield
  {journal} {\bibinfo  {journal} {Physical Review A}\ }\textbf {\bibinfo
  {volume} {45}},\ \bibinfo {pages} {7184} (\bibinfo {year}
  {1992})}\BibitemShut {NoStop}%
\bibitem [{\citenamefont {Debye}(1929)}]{debye29PolarMolec}%
  \BibitemOpen
  \bibfield  {author} {\bibinfo {author} {\bibfnamefont {P.}~\bibnamefont
  {Debye}},\ }\href@noop {} {\emph {\bibinfo {title} {Polar Molecules}}}\
  (\bibinfo  {publisher} {Dover},\ \bibinfo {address} {New York},\ \bibinfo
  {year} {1929})\BibitemShut {NoStop}%
\bibitem [{\citenamefont {Das}\ \emph {et~al.}(2015)\citenamefont {Das},
  \citenamefont {Garg}, \citenamefont {Campbell}, \citenamefont {Howse},
  \citenamefont {Sen}, \citenamefont {Velegol}, \citenamefont {Golestanian},\
  and\ \citenamefont {Ebbens}}]{Das2015BoundariesSpheres}%
  \BibitemOpen
  \bibfield  {author} {\bibinfo {author} {\bibfnamefont {S.}~\bibnamefont
  {Das}}, \bibinfo {author} {\bibfnamefont {A.}~\bibnamefont {Garg}}, \bibinfo
  {author} {\bibfnamefont {A.~I.}\ \bibnamefont {Campbell}}, \bibinfo {author}
  {\bibfnamefont {J.}~\bibnamefont {Howse}}, \bibinfo {author} {\bibfnamefont
  {A.}~\bibnamefont {Sen}}, \bibinfo {author} {\bibfnamefont {D.}~\bibnamefont
  {Velegol}}, \bibinfo {author} {\bibfnamefont {R.}~\bibnamefont
  {Golestanian}}, \ and\ \bibinfo {author} {\bibfnamefont {S.~J.}\ \bibnamefont
  {Ebbens}},\ }\href {https://www.nature.com/articles/ncomms9999#citeas}
  {\bibfield  {journal} {\bibinfo  {journal} {Nature Communications}\ }\textbf
  {\bibinfo {volume} {6}},\ \bibinfo {pages} {8999} (\bibinfo {year}
  {2015})}\BibitemShut {NoStop}%
\bibitem [{\citenamefont {Simmchen}\ \emph {et~al.}(2016)\citenamefont
  {Simmchen}, \citenamefont {Katuri}, \citenamefont {Uspal}, \citenamefont
  {Popescu}, \citenamefont {Tasinkevych},\ and\ \citenamefont
  {S{\'{a}}nchez}}]{Simmchen2016TopographicalMicroswimmers}%
  \BibitemOpen
  \bibfield  {author} {\bibinfo {author} {\bibfnamefont {J.}~\bibnamefont
  {Simmchen}}, \bibinfo {author} {\bibfnamefont {J.}~\bibnamefont {Katuri}},
  \bibinfo {author} {\bibfnamefont {W.~E.}\ \bibnamefont {Uspal}}, \bibinfo
  {author} {\bibfnamefont {M.~N.}\ \bibnamefont {Popescu}}, \bibinfo {author}
  {\bibfnamefont {M.}~\bibnamefont {Tasinkevych}}, \ and\ \bibinfo {author}
  {\bibfnamefont {S.}~\bibnamefont {S{\'{a}}nchez}},\ }\href
  {https://www.nature.com/articles/ncomms10598} {\bibfield  {journal} {\bibinfo
   {journal} {Nature Communications}\ }\textbf {\bibinfo {volume} {7}},\
  \bibinfo {pages} {10598} (\bibinfo {year} {2016})}\BibitemShut {NoStop}%
\bibitem [{\citenamefont {Holubec}\ \emph {et~al.}(2019)\citenamefont
  {Holubec}, \citenamefont {Kroy},\ and\ \citenamefont
  {Steffenoni}}]{holubec2019FPESolver}%
  \BibitemOpen
  \bibfield  {author} {\bibinfo {author} {\bibfnamefont {V.}~\bibnamefont
  {Holubec}}, \bibinfo {author} {\bibfnamefont {K.}~\bibnamefont {Kroy}}, \
  and\ \bibinfo {author} {\bibfnamefont {S.}~\bibnamefont {Steffenoni}},\
  }\href {https://doi.org/10.1103/PhysRevE.99.032117} {\bibfield  {journal}
  {\bibinfo  {journal} {Phys. Rev. E}\ }\textbf {\bibinfo {volume} {99}},\
  \bibinfo {pages} {032117} (\bibinfo {year} {2019})}\BibitemShut {NoStop}%
\bibitem [{\citenamefont {Solon}\ \emph {et~al.}(2015)\citenamefont {Solon},
  \citenamefont {Stenhammar}, \citenamefont {Wittkowski}, \citenamefont
  {Kardar}, \citenamefont {Kafri}, \citenamefont {Cates},\ and\ \citenamefont
  {Tailleur}}]{Solon2015PressureSpheres}%
  \BibitemOpen
  \bibfield  {author} {\bibinfo {author} {\bibfnamefont {A.~P.}\ \bibnamefont
  {Solon}}, \bibinfo {author} {\bibfnamefont {J.}~\bibnamefont {Stenhammar}},
  \bibinfo {author} {\bibfnamefont {R.}~\bibnamefont {Wittkowski}}, \bibinfo
  {author} {\bibfnamefont {M.}~\bibnamefont {Kardar}}, \bibinfo {author}
  {\bibfnamefont {Y.}~\bibnamefont {Kafri}}, \bibinfo {author} {\bibfnamefont
  {M.~E.}\ \bibnamefont {Cates}}, \ and\ \bibinfo {author} {\bibfnamefont
  {J.}~\bibnamefont {Tailleur}},\ }\href {\doibase
  10.1103/PhysRevLett.114.198301} {\bibfield  {journal} {\bibinfo  {journal}
  {Physical Review Letters}\ }\textbf {\bibinfo {volume} {114}},\ \bibinfo
  {pages} {198301} (\bibinfo {year} {2015})}\BibitemShut {NoStop}%
\bibitem [{\citenamefont {Solon}\ \emph
  {et~al.}(2018{\natexlab{b}})\citenamefont {Solon}, \citenamefont
  {Stenhammar}, \citenamefont {Cates}, \citenamefont {Kafri},\ and\
  \citenamefont {Tailleur}}]{solon2018GenTDofMIPS}%
  \BibitemOpen
  \bibfield  {author} {\bibinfo {author} {\bibfnamefont {A.~P.}\ \bibnamefont
  {Solon}}, \bibinfo {author} {\bibfnamefont {J.}~\bibnamefont {Stenhammar}},
  \bibinfo {author} {\bibfnamefont {M.~E.}\ \bibnamefont {Cates}}, \bibinfo
  {author} {\bibfnamefont {Y.}~\bibnamefont {Kafri}}, \ and\ \bibinfo {author}
  {\bibfnamefont {J.}~\bibnamefont {Tailleur}},\ }\href {\doibase
  10.1088/1367-2630/aaccdd} {\bibfield  {journal} {\bibinfo  {journal} {New
  Journal of Physics}\ }\textbf {\bibinfo {volume} {20}},\ \bibinfo {pages}
  {075001} (\bibinfo {year} {2018}{\natexlab{b}})}\BibitemShut {NoStop}%
\bibitem [{\citenamefont {Caprini}\ \emph {et~al.}(2020)\citenamefont
  {Caprini}, \citenamefont {Marini Bettolo~Marconi},\ and\ \citenamefont
  {Puglisi}}]{caprini2020velocityAlignment}%
  \BibitemOpen
  \bibfield  {author} {\bibinfo {author} {\bibfnamefont {L.}~\bibnamefont
  {Caprini}}, \bibinfo {author} {\bibfnamefont {U.}~\bibnamefont {Marini
  Bettolo~Marconi}}, \ and\ \bibinfo {author} {\bibfnamefont {A.}~\bibnamefont
  {Puglisi}},\ }\href {https://doi.org/10.1103/PhysRevLett.124.078001}
  {\bibfield  {journal} {\bibinfo  {journal} {Physical Review Letters}\
  }\textbf {\bibinfo {volume} {124}},\ \bibinfo {pages} {078001} (\bibinfo
  {year} {2020})}\BibitemShut {NoStop}%
\end{thebibliography}%


%
\end{document}